\documentclass[preprints,article,accept,moreauthors,pdflatex]{mdpi} 

\firstpage{1} 
\pubvolume{1}
\issuenum{1}
\articlenumber{0}
\pubyear{2021}
\copyrightyear{2021}
\datereceived{30 July 2021} 
\dateaccepted{1 September 2021} 
\datepublished{} 
\hreflink{https://doi.org/} 

\pdfoutput=1

\Title{Inclination estimates from off-axis GRB afterglow modelling}

\TitleCitation{Inclination estimates from off-axis GRB afterglow modelling}

\Author{Gavin P. Lamb* $^{1}$\orcidA{}, Joseph J. Fern\'andez $^{2}$\orcidB{}, Fergus Hayes $^{3}$\orcidC{}, Albert K. H. Kong $^{4}$\orcidD{}, En-Tzu Lin $^{4}$\orcidE{}, Nial R. Tanvir $^{1}$\orcidF{}, Martin Hendry $^{3}$\orcidG{}, Ik Siong Heng $^{3}$\orcidH{}, Surojit Saha $^{4}$\orcidJ{}, John Veitch $^{3}$\orcidI{}}

\AuthorNames{Gavin P. Lamb, Joseph J. Fern\'andez, Fergus Hayes, Albert K. H. Kong, En-Tzu Lin, Nial R. Tanvir}

\AuthorCitation{Lamb, G. P.; ...}

\address{%
$^{1}$ \quad School of Physics and Astronomy, University of Leicester, Leicester LE1 7RH, UK \\
$^{2}$ \quad Astrophysics Research Institute, Liverpool John Moores University, Liverpool L3 5RF, UK\\
$^{3}$ \quad SUPA, School of Physics and Astronomy, University of Glasgow, Glasgow G12 8QQ, UK\\
$^{4}$ \quad Institute of Astronomy, National Tsing Hua University, Hsinchu, Taiwan, ROC}

\corres{Correspondence: gpl6@le.ac.uk}

\abstract{For gravitational wave (GW) detected neutron star mergers, one of the leading candidates for electromagnetic (EM) counterparts is the afterglow from an ultra-relativistic jet. Where this afterglow is observed, it will likely be viewed off-axis, such as the afterglow following GW170817/GRB 170817A. The temporal behaviour of an off-axis observed GRB afterglow can be used to reveal the lateral jet structure, and statistical model fits can put constraints on the various model free-parameters. Amongst these parameters is the inclination of the system to the line of sight. Along with the GW detection, the afterglow modelling provides the best constraint on the inclination to the line-of-sight and can improve the estimates of cosmological parameters e.g. the Hubble constant, from GW-EM events. However, modelling of the afterglow depends on the assumed jet structure and, often overlooked, the effects of lateral spreading. Here we show how the inclusion of lateral spreading in the afterglow models can affect the estimated inclination of GW-EM events.}


\begin{document}

\section{Introduction}

Electromagnetic (EM) counterparts to gravitational wave (GW) detected events (GW-EM) are amongst the most promising multi-messenger astronomy sources.
Where GWs are detected from neutron star mergers, the potential EM counterparts include macro-, kilo-, or merger-nova, Gamma-Ray Bursts (GRBs), and GRB afterglows \citep{nakar2020}.
For a GW detection of a neutron star merger, the EM afterglows that are typically associated with GRBs offer a unique opportunity to probe these systems via a new trigger, the GW signal.
As GRBs are highly beamed, they are typically seen at very small system inclinations to the line-of-sight, however, a GW triggered GRB producing system will likely be seen at a much higher inclination.
In such cases the prompt GRB will likely be absent, or faint, and the afterglow will appear unique when compared to the cosmological population.
GRB afterglows as GW-EMs will probe the structure of the outflows and jets that produce a fraction of the cosmological population of GRBs \citep{lamb2017, lazzati2017, jin2018, kathirgamaraju2018}.

GW-EM counterparts can be used to maximise the science returns from GW astronomy, and amongst their possible uses, the EM counterparts can help put constraints on various cosmological parameters measurable via GW astronomy.
For GRB afterglow GW-EM counterparts, the rise index of the afterglow can give information about the lateral structure of the jet, and by assuming a fixed jet profile or structure, this rise index can be used to put tighter constraints on the inclination angle to the line-of-sight, $\iota$, for an observed afterglow \citep[e.g.][]{wang2021}.
By improving the constraints on $\iota$, the luminosity distance as measured by GWs, and a redshift from a host galaxy can be used to give tighter constraints on the Hubble parameter, $H_0$ \citep[e.g.][]{hotokezaka2019, mastrogiovanni2020a}.
Here we ask the question, how does the inclusion or omission of lateral spreading within the GRB afterglow calculation affect the inferred inclination angle of an observed system?

The lateral spreading of the blastwave that produces a GRB afterglow has been described via hydrodynamic simulations \cite[e.g.][]{vaneerten2012} and analytically \cite[e.g.][]{granot2012}, however, lateral spreading is often neglected in simple afterglow modelling.
The effects of lateral spreading on radio images of GW-EM afterglows has been investigated by \cite{fernandez2021}, and here we show how lateral spreading affects the shape of the afterglow lightcurve with inclination.
For each case we show results without lateral spreading, and with lateral spreading at the sound-speed \citep[e.g.][]{huang2000, lamb2021}, which can be considered a physical maximum on the effects of spreading.

In \S \ref{sec:method} we describe our methods for generating afterglows and fitting these models to data.
In \S \ref{sec:results} we show the results of our models and fits, and in \S \ref{sec:disc} we discuss our results and interpret these in terms of GW-EM observations.
Our conclusions are presented in \S \ref{sec:conc}.

\section{Method}\label{sec:method}

We generate afterglow lightcurves for a variety of fiducial jet structure models defined by unique energy and Lorentz factor profiles with polar angle of the outflow, or jet.
The synchrotron emission from the shocked electrons within the forward shock of a relativistic blastwave and the effects of synchrotron self absorption, which is particularly important at very low frequencies, is calculated and the emission from the equal arrival time surface is summed to give the total flux with observer time.
We use the method developed and utilised for the prediction of structured jet counterparts to gravitational wave detected neutron star mergers and the modelling of the afterglow to GRB\,170817A \citep[see][for details]{lamb2017, lamb2018a, lamb2018b, lyman2018, lamb2019a, lamb2019b, lamb2020, lamb2021}, both with and without the inclusion of lateral spreading.
{Generally jet structures are divided by on-axis core emission described by a core energy and Lorentz factor along the jet axis, and lower energy off-axis emission at angles greater than a core angle $\theta_c$.}

The jet structure profiles are described as:
\begin{itemize}
    \item[\bf{TH}] Top-hat: a uniform in { isotropic-equivalent kinetic} energy, $E$, and Lorentz factor, $\Gamma$, with angle until a sharp edge at the jet opening angle, $\theta_j \equiv \theta_c$, where the energy goes to zero.
    \item[\bf{G }] Gaussian: a jet profile described by a Gaussian function with $E(\theta) = E(\theta=0){\rm exp}[-\theta^2/\theta_c^2]$, and $\Gamma(\theta) = (\Gamma(\theta=0)-1){\rm exp}[-\theta^2/(2\theta_c^2)]+1$.
    \item[\bf{2C}] Two-component: a top-hat jet surrounded by {a wider uniform region of lower energy} with $E(\theta>\theta_c) = 0.1 E(\theta=0)$, and $\Gamma(\theta>\theta_c) = 5$.
    \item[\bf{PL}] Powerlaw: a top-hat jet surrounded by {a region where the energy and Lorentz factor declines with increasing angle as a powerlaw}, $E(\theta>\theta_c) = E(\theta=0)[\theta/\theta_c]^{-2}$, and $\Gamma(\theta>\theta_c) = (\Gamma(\theta=0)-1)[\theta/\theta_c]^{-2}+1$.
\end{itemize}
All jets have a maximum opening angle where energy goes to zero at $\theta_j=0.35$ except the top-hat profile, which goes to zero at $\theta_c$.

The lateral spreading prescription we use assumes a sideways expansion i.e. perpendicular to the radial direction of the jet, at the local sound-speed, $c_s$.
The sound-speed is found at each step in the solution for the radial dynamics of the blastwave \citep[for details see][]{lamb2018b,lamb2021}, and used to calculated the change in the solid angle of the jet as it spreads.
Where lateral spreading is not included, the jet is assumed to maintain the same solid angle throughout its lifetime i.e. it expands radially within a conic volume described by the opening angle of the jet, $\theta_j$.
These two limits represent the physical extremes of the possible lateral spreading for a relativistic jet; no lateral spread, or sound-speed lateral spreading.

For our fiducial jet models we set identical afterglow parameters so as to better highlight the effects of including lateral spreading on the lightcurves.
Our fixed parameters are:
isotropic{-equivalent kinetic} energy { along the jet central axis}, $E(\theta=0) = 10^{51}$ erg,
initial Lorentz factor { along the jet central axis}, $\Gamma(\theta=0) = 100$,
ambient particle number density, $n=0.1$\,cm$^{-3}$,
microphysical parameters, $\varepsilon_B=\varepsilon_e^2=0.01$, and the shocked electron index, $p=2.5$,
and we fix the core angle, $\theta_c=0.1$\,rad.

The afterglows for the four jet structure models, each with and without lateral spreading are calculated at a radio frequency of 3\,GHz.
We find the temporal index, $\alpha$, with observer time using,
\begin{equation}
    \alpha = \frac{{\rm d} \log F}{{\rm d} \log t},
    \label{eq:alpha}
\end{equation}
where, $F$ is the flux density, and $t$ is the observed time.
{ We use} Richardson's extrapolation { to find the first derivative of the $\log F$} lightcurve { with $\log t$} \cite{richardson1911ix}.

The rise index pre-afterglow peak flux is of particular interest, as this has been used to put constraints on the line-of-sight inclination, $\iota$, \citep{wang2021}.
For each jet structure model we show how the $\alpha$ changes with inclination at $0.3 t_p$, where $t_p$ is the observed peak flux time.
We find $t_p$ and $\alpha(0.3t_p)$ for all four models both with and without lateral spreading.

A real test of the effects of lateral spreading on the parameter estimation for GW-EM, GRB afterglows is demonstrated by fitting, via Markov Chain Monte Carlo (MCMC), the four jet structure models with/without spreading to the data from GRB\,170817A -- we use the latest { radio through X-ray frequency} data as listed in \cite{troja2021}, \cite{balasubramanian2021},\cite{makhathini2020}, and \cite{fong2019}.
For the fits to GRB\,170817A, we fix $p=2.15$ \cite[e.g.,][]{troja2018,lamb2019a} and allow all other parameters to vary using a flat prior in each case as defined in Table \ref{tab1}.
We use \texttt{emcee} for our MCMC \citep{foreman2013}.

\section{Results}\label{sec:results}

\begin{figure}
    \includegraphics[width=\columnwidth]{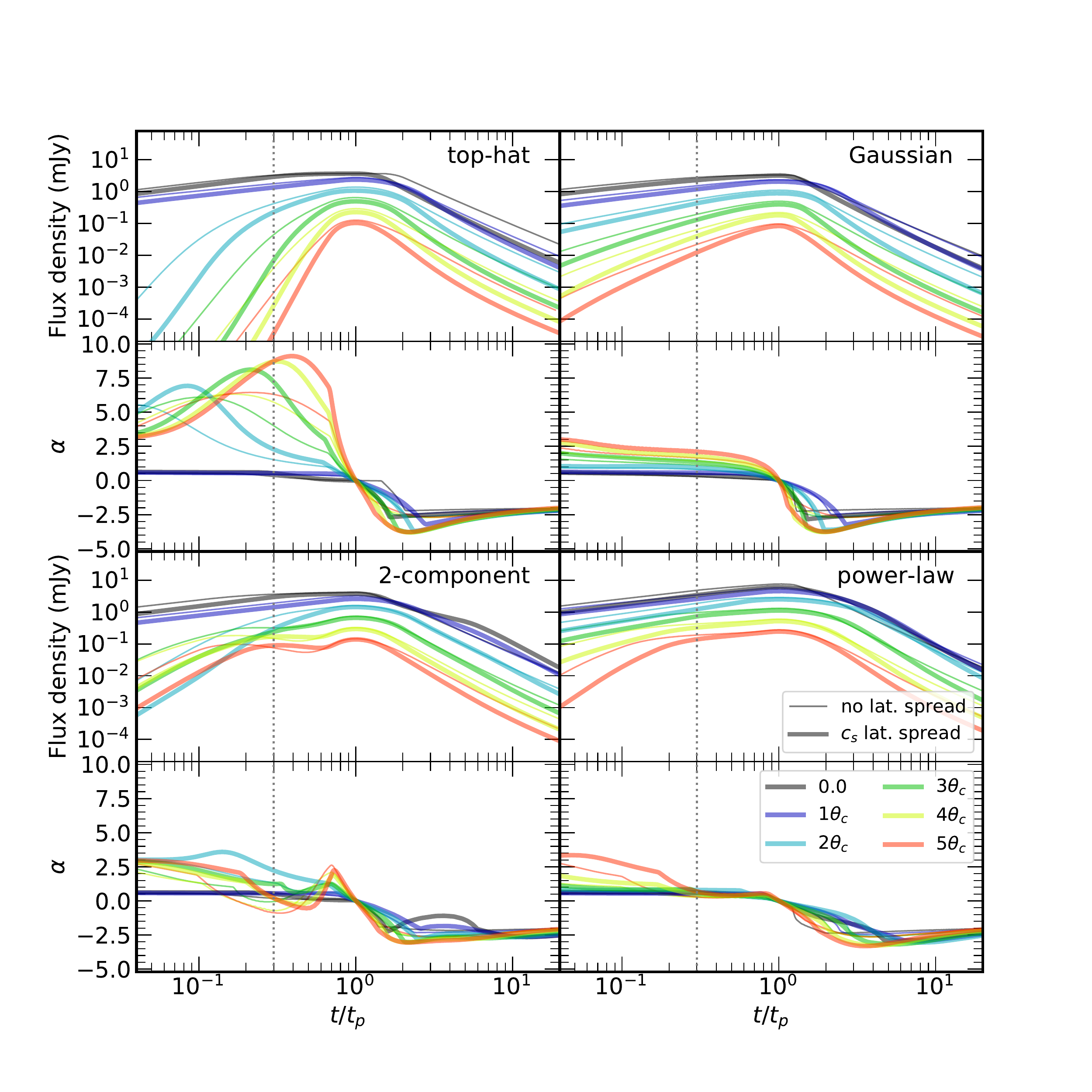}
    \caption{Lightcurves and temporal index, $\alpha$, evolution relative to the lightcurve peak time for four jet structure profile models.
    Different line colours indicate the line-of-sight angle from the jet central axis.
    Line width indicates whether lateral spreading is included or not; a thin line without spreading, and a thick line with sound speed spreading.
    The vertical grey dotted line indicates $t=0.3 t_p$, and is the time used to show the change in $\alpha$ with inclination.}
    \label{fig:lc-alpha}
\end{figure}

Figure \ref{fig:lc-alpha} shows the afterglow lightcurves at 3 GHz for jets with our fiducial parameters and viewed at inclination angles $0.0\leq\iota\leq0.5$\,rad, or with our fixed $\theta_c$ equivalent to 0 -- 5$\theta_c$.
Each of the four panels shows the lightcurve, $F(t)$ (top half) and $\alpha(t)$ (bottom half) at one of six inclinations.
The observer time is normalised in each case to the afterglow peak time, $t_p$, and a vertical dotted line indicates $0.3t_p$.

\begin{figure}
    \includegraphics[width=\columnwidth]{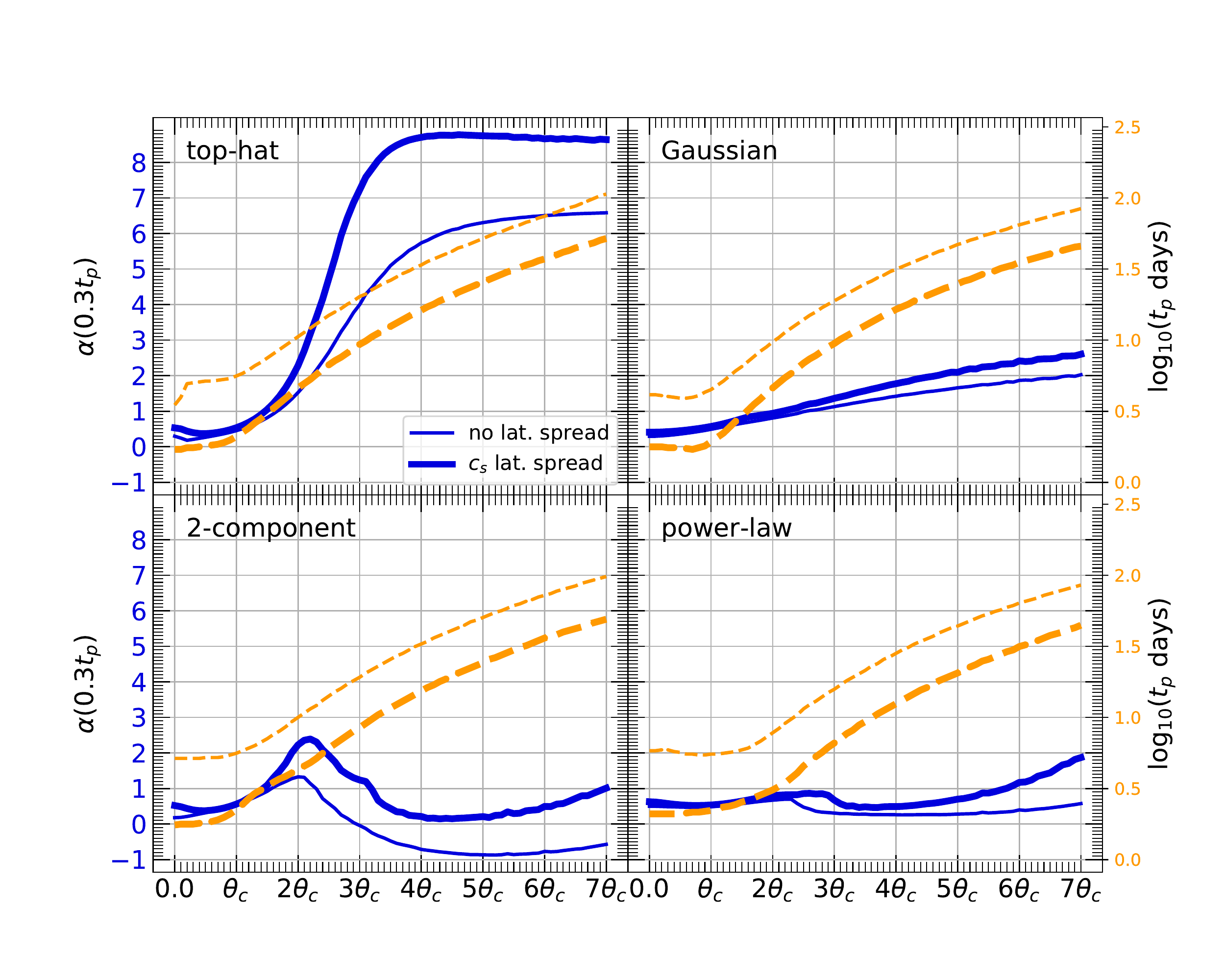}
    \caption{The temporal index at $0.3\times$ the observed lightcurve peak time at increasing inclinations (blue lines, left axis), and the observed lightcurve peak time for our model parameters for the cases with and without lateral spreading (orange, right axis).
    { The x-axis shows the inclination as a factor of the core angle, $\theta_c=0.1$\,rad.}
    The lateral spreading model assumes a maximum sound speed sideways expansion and so should be considered an upper limit on the effect.}
    \label{fig:alpha-0.3tp}
\end{figure}

Figure \ref{fig:alpha-0.3tp} shows how $\alpha(0.3t_p)$ changes with inclination for each of the four jet structures, blue lines, both with (thick solid line) and without (thin solid line) lateral spreading.
The afterglow observed peak time for our parameters with inclination is shown with an orange dashed line.
For all inclination angles $\iota>\theta_c$, the lightcurve from the sound-speed lateral spreading case rises with a larger index than the case without lateral spreading.
Similarly, we note that where the parameters of the jet are identical, the inclusion of maximal lateral spreading acts to reduce the observed peak flux time in each case by a factor $\sim2$.
In the case of the two-component model, the afterglow can exhibit multiple peaks and, at our reference time where lateral spreading is not included, is declining for $\iota>3\theta_c$.
The equivalent lightcurves, that include lateral spreading, show a plateau or very shallow rise index for some off-axis observers at $t=0.3t_p$.
The change in rise index for the more smooth edged jet structure profiles i.e. Gaussian and powerlaw, are less extreme at $t=0.3t_p$ with $\alpha\lesssim2.5$ when compared to the classic, top-hat profile, with a rapid rise to peak of $\alpha\sim6$ -- $8$.

\begin{figure}
    \includegraphics[width=\columnwidth]{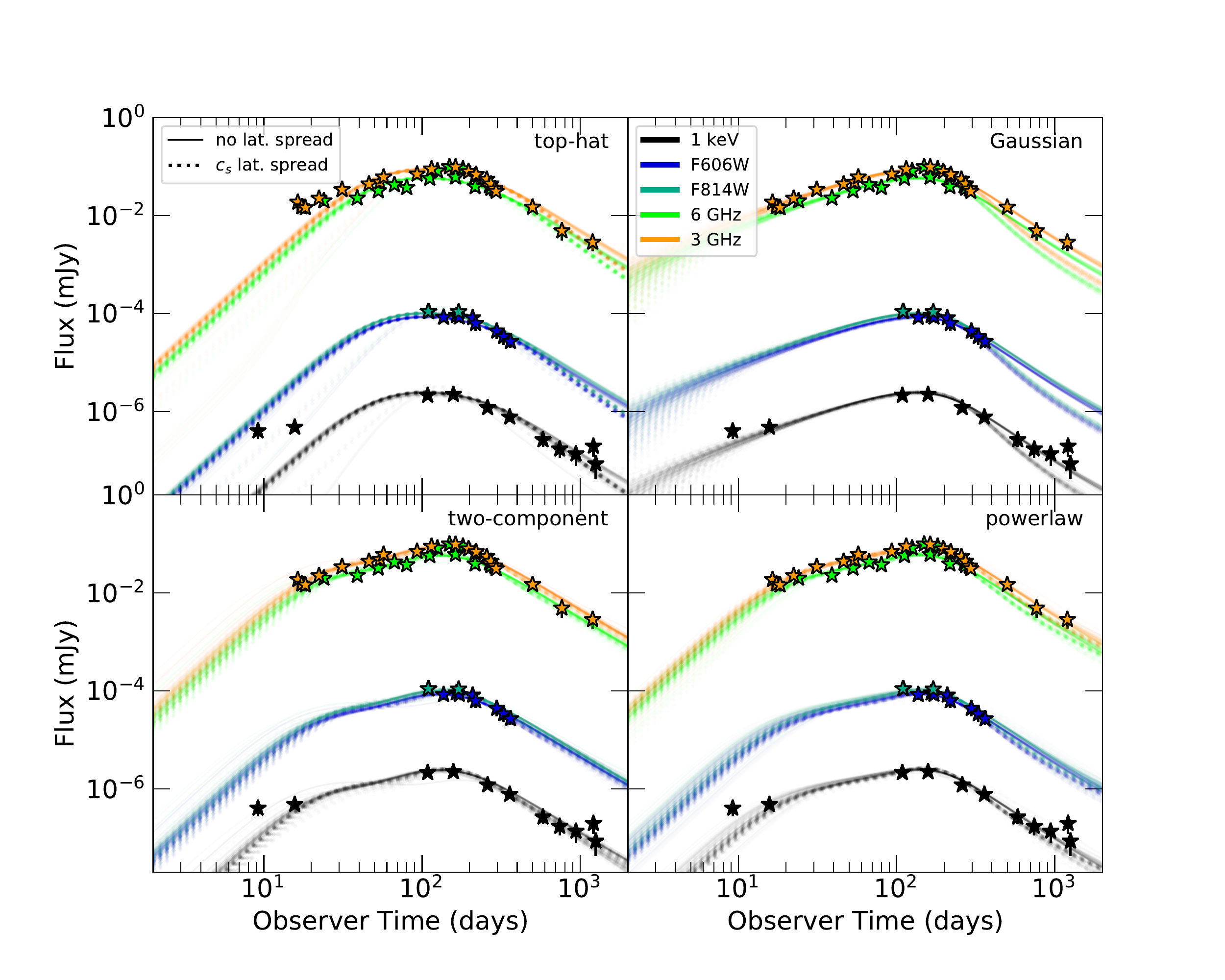}
    \caption{Afterglow lightcurve fits to the late-time afterglow to GW170817 { at radio (3 and 6 GHz), infrared and optical ({\it Hubble Space Telescope} filters F814W and F606W), and X-ray (1 keV)}. 
    We use a Markov Chain Monte Carlo (MCMC) and plot a random sample of 100 lightcurves drawn from the posterior distributions.
    The four jet structures are top-hat (top left), Gaussian (top right), two-component (bottom left), and a powerlaw (bottom right) jet structure profiles. 
    Each panel includes lightcurves with and without lateral spreading (dotted and solid lines respectively). Where the models with and without lateral spreading diverge we find that, typically, the model where lateral spreading is included will have a steeper decline immediately after the lightcurve peak -- most obvious in the case of the Gaussian jet structure profile.
    }
    \label{fig:GW170817}
\end{figure}

Each of these jet structure models can be fit to data from a GW-EM GRB afterglow counterpart to see how these effects alter the inferred parameters.
We fit via MCMC the four jet { structure} profiles with and without lateral spreading.
One hundred random posterior sample lightcurve draws
of the fits to the broadband afterglow data of GRB\,170817A are shown in Figure \ref{fig:GW170817}.
The dotted lines show the cases with lateral spreading, whereas the solid lines show the case without spreading.
From this figure, it is clear that, generally, it is the very late post-peak lightcurve that is most sensitive to the inclusion of lateral spreading, as noted in \cite{lamb2018b, troja2018}.
We additionally note that the top-hat jet profile { does not} 
satisfactorily describe the early radio and X-ray data, however, this may be due to the omission of a non-zero shell thickness in the afterglow model { or a more realistic early relaxation of the jet structure due to lateral spreading} \citep[see][]{gill2019}
although, the leading explanation for the early afterglow data is the contribution from the wider components of a structured, relativistic jet { or the fastest components of an accompanying cocoon surrounding the jet} \citep[][etc.]{lazzati2018, lyman2018, troja2018, lamb2018a, margutti2018, resmi2018}, alternatively, energy injection into the jetted blastwave can account for the observed afterglow and the system inclination for a top-hat jet model \citep{lamb2020}. 

\begin{specialtable} 
\caption{The prior range and central values for the fit parameters from the posterior distributions for each of the jet structure models with (w) and without (wo) lateral spreading for lightcurve fits to the broadband afterglow data to GW170817/GRB\,170817A. {The posterior upper and lower bounds are represented by the 16th and 84th percentile and the central value by the median.}}
\resizebox{\columnwidth}{!}{\begin{tabular}{ccccccccc}
\toprule
{} & {} & $\log\varepsilon_B$ & $\log\varepsilon_e$ & $\log n$ & $\theta_c$ & $\log E(\theta=0)$ & $\Gamma(\theta=0)$ & $\cos\iota=\mu$
\\
\multicolumn{2}{c} {units} & {} & {} & [$\log$ cm$^{-3}$] & [rad] & [$\log$ erg] & {} & {} \\
\midrule
\multicolumn{2}{c} {Prior}    & $-6$ -- $-0.3$ & $-6$ -- $-0.3$ & $-6$ -- $2$ & $0.01$ -- $0.1$ & $48$ -- $55$ & $6$ -- $600$ & $\cos(2\theta_c $-- $10\theta_c)$ \\
\\
\bf{TH} & wo & 
$ -2.9 ^{+ 1.6 }_{- 1.7 }$  & 
$ -2.5 ^{+ 1.1 }_{- 1.6 }$  & 
$ -3.4 ^{+ 1.1 }_{- 0.9 }$  & 
$ 0.08 ^{+ 0.01 }_{- 0.01 }$  & 
$ 52.9 ^{+ 1.4 }_{- 1.1 }$  & 
$ 12.1 ^{+ 6.1 }_{- 2.6 }$  & 
$ 0.980 ^{+ 0.006 }_{- 0.009 }$ \\
        & w  & 
$ -2.8 ^{+ 1.6 }_{- 1.9 }$  & 
$ -2.4 ^{+ 1.3 }_{- 1.4 }$  & 
$ -3.9 ^{+ 1.1 }_{- 0.9 }$  & 
$ 0.08 ^{+ 0.01 }_{- 0.02 }$  & 
$ 53.1 ^{+ 1.2 }_{- 1.0 }$  & 
$ 10.7 ^{+ 2.9 }_{- 1.8 }$  & 
$ 0.984 ^{+ 0.006 }_{- 0.004 }$ \\
\bf{G}  & wo & 
$ -3.8 ^{+ 1.8 }_{- 1.4 }$  & 
$ -2.0 ^{+ 0.8 }_{- 1.4 }$  & 
$ -2.5 ^{+ 1.3 }_{- 1.1 }$  & 
$ 0.07 ^{+ 0.02 }_{- 0.02 }$  & 
$ 53.0 ^{+ 1.1 }_{- 0.9 }$  & 
$ 365.1 ^{+ 137.4 }_{- 153.6 }$  & 
$ 0.958 ^{+ 0.017 }_{- 0.021 }$ \\
        & w  & 
$ -2.9 ^{+ 1.6 }_{- 2.0 }$  & 
$ -2.2 ^{+ 1.0 }_{- 1.2 }$  & 
$ -3.0 ^{+ 1.1 }_{- 0.9 }$  & 
$ 0.09 ^{+ 0.01 }_{- 0.01 }$  & 
$ 53.1 ^{+ 1.1 }_{- 1.1 }$  & 
$ 181.0 ^{+ 112.5 }_{- 84.7 }$  & 
$ 0.947 ^{+ 0.011 }_{- 0.007 }$ \\
\bf{2C} & wo & 
$ -2.6 ^{+ 1.5 }_{- 1.6 }$  & 
$ -2.3 ^{+ 1.1 }_{- 1.3 }$  & 
$ -3.5 ^{+ 1.5 }_{- 0.7 }$  & 
$ 0.06 ^{+ 0.01 }_{- 0.01 }$  & 
$ 52.4 ^{+ 1.4 }_{- 0.6 }$  & 
$ 316.1 ^{+ 152.9 }_{- 160.7 }$  & 
$ 0.974 ^{+ 0.001 }_{- 0.003 }$ \\
        & w  & 
$ -1.9 ^{+ 1.1 }_{- 2.0 }$  & 
$ -1.5 ^{+ 0.6 }_{- 1.1 }$  & 
$ -3.6 ^{+ 0.9 }_{- 0.8 }$  & 
$ 0.08 ^{+ 0.01 }_{- 0.01 }$  & 
$ 52.0 ^{+ 1.2 }_{- 0.5 }$  & 
$ 301.2 ^{+ 85.4 }_{- 129.4 }$  & 
$ 0.949 ^{+ 0.007 }_{- 0.089 }$ \\
\bf{PL} & wo & 
$ -2.2 ^{+ 1.3 }_{- 1.7 }$  & 
$ -0.9 ^{+ 0.4 }_{- 1.5 }$  & 
$ -3.5 ^{+ 1.2 }_{- 0.9 }$  & 
$ 0.03 ^{+ 0.01 }_{- 0.01 }$  & 
$ 51.7 ^{+ 1.5 }_{- 1.0 }$  & 
$ 105.6 ^{+ 43.9 }_{- 33.8 }$  & 
$ 0.957 ^{+ 0.021 }_{- 0.032 }$ \\
        & w  & 
$ -3.4 ^{+ 1.9 }_{- 1.7 }$  & 
$ -1.4 ^{+ 0.8 }_{- 1.4 }$  & 
$ -1.8 ^{+ 1.5 }_{- 1.5 }$  & 
$ 0.06 ^{+ 0.01 }_{- 0.01 }$  & 
$ 52.3 ^{+ 1.4 }_{- 0.9 }$  & 
$ 105.0 ^{+ 43.9 }_{- 26.2 }$  & 
$ 0.839 ^{+ 0.050 }_{- 0.018 }$ \\
\bottomrule
\end{tabular}}
\label{tab1}
\end{specialtable}

\begin{figure}
    \includegraphics[width=\columnwidth]{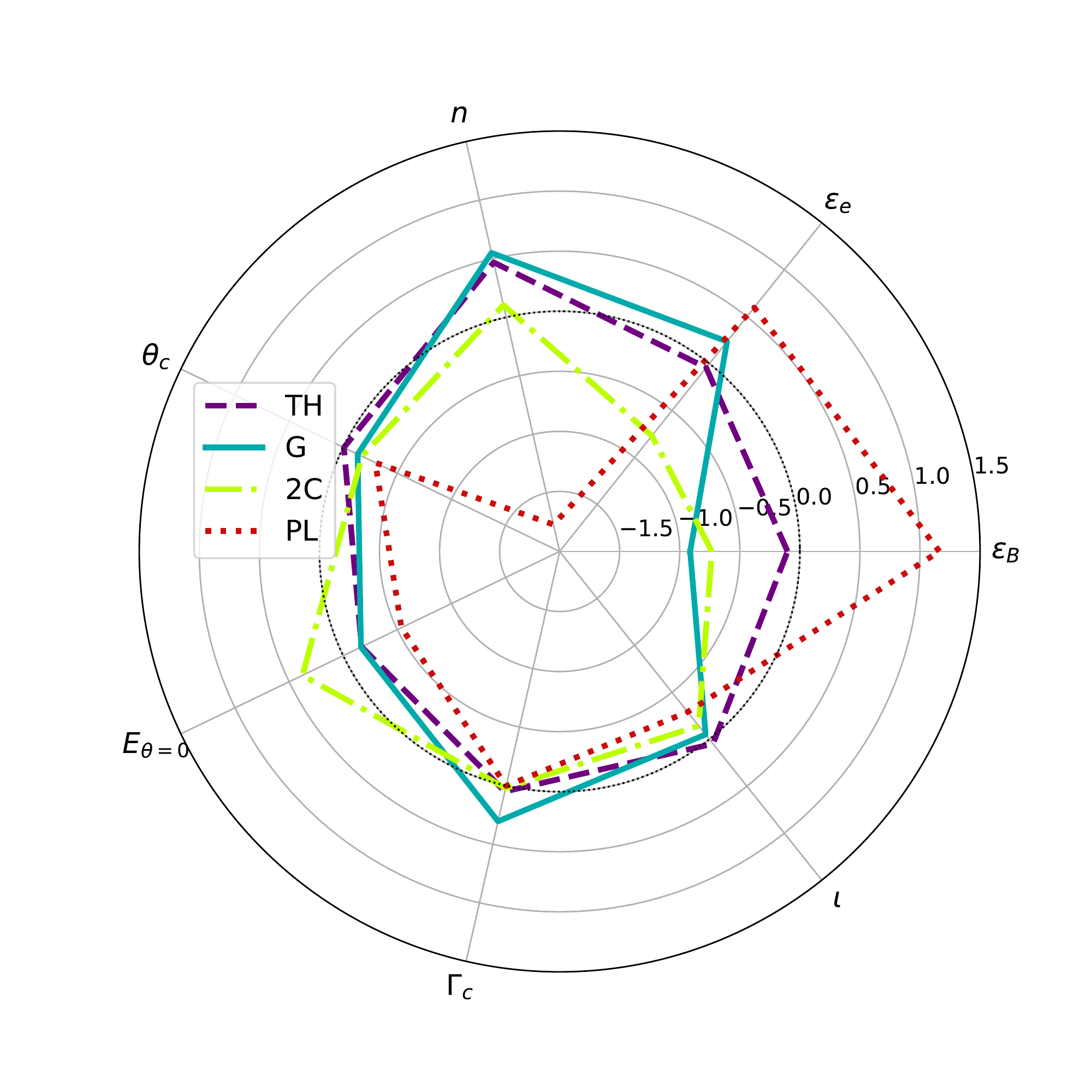}
    \caption{A polar plot showing the log ratio values for each parameter from a fixed jet structure model fit to GRB\,170817A data via MCMC, \textbf{without and with lateral spreading.}
    Each parameter distribution's \textbf{median} and 16th and 84th percentile limits are listed in table \ref{tab1} for all models.
    The radial arm shows the $\log_{10}({\rm wo}/{\rm w})$ of the central posterior distribution values for the labelled parameters, where wo is without lateral spreading and w is with maximal lateral spreading at the sound speed.
    An equal ratio of one is indicated with a black dotted line.
    The top-hat structure is shown in purple with a dashed line, a Gaussian jet structure is shown in aqua with a solid line, a two-component jet is shown in green with a dot-dashed line, and a powerlaw structure is shown in red with a dotted line.}
    \label{fig:ratios}
\end{figure}

\begin{figure}
    \includegraphics[width=\columnwidth]{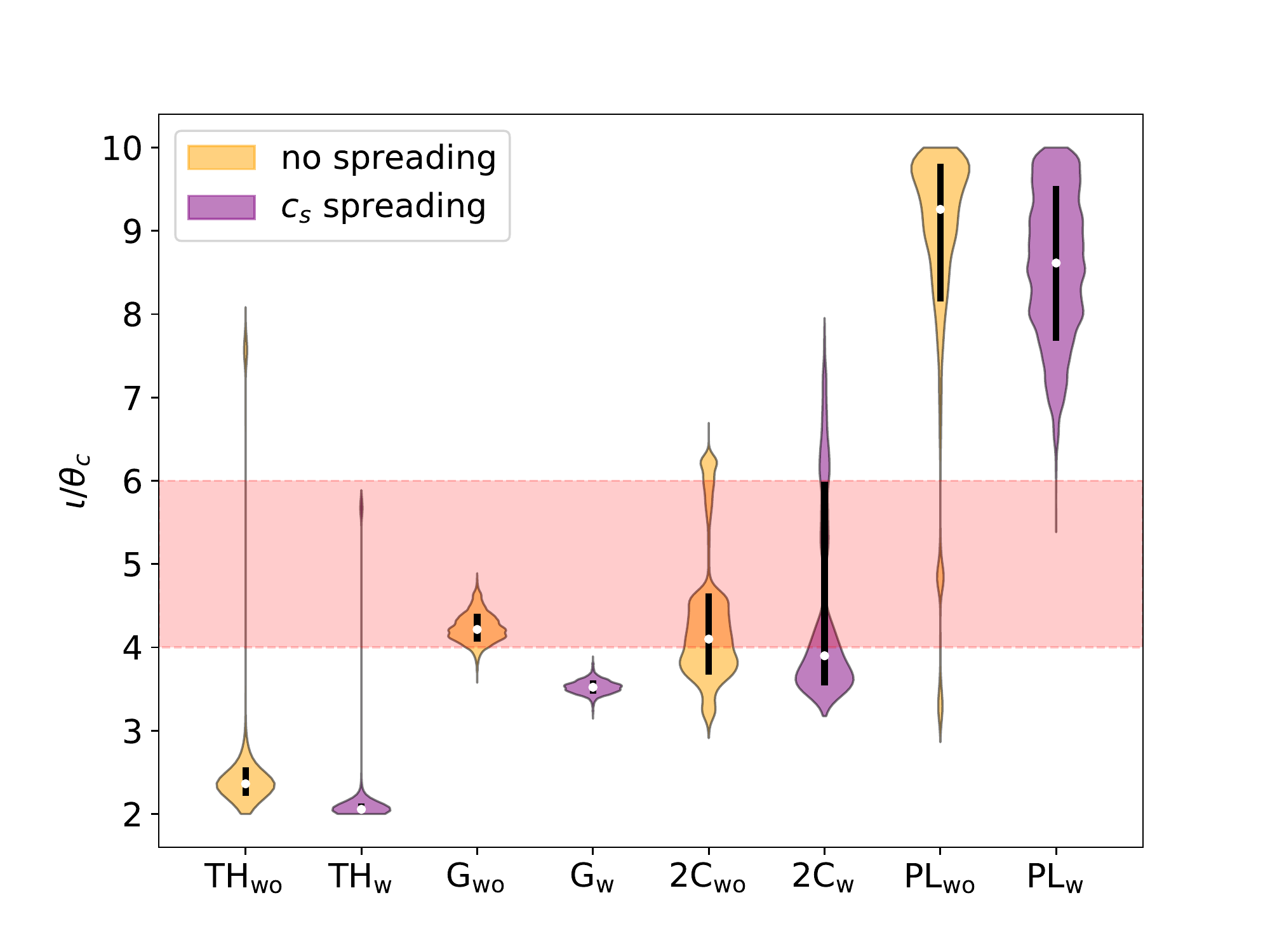}
    \caption{The posterior distribution from MCMC fits of our four fiducial jet structure models to GW170817 late-time afterglow data with no lateral spreading (orange) and maximal lateral spreading (purple). The white dot and the thick black lines show the central and the 16th and 84th percentile values for the distribution. { The horizontal width is indicative of the probability distribution in $\iota/\theta_c$ for each model.} The expected ratio for consistency with VLBI observations of superluminal motion are  $4\lesssim\iota/\theta_c\lesssim6$ \citep{nakar2021}, red band, and where values of $\iota/\theta_c<6$ can give a reasonable approximation for the real observation angle when inferred from superluminal motion measurements via the point approximation \citep{fernandez2021}.}
    \label{fig:violin}
\end{figure}

The prior range and the central values for each parameter in the posterior distributions for these lightcurves is listed in Table \ref{tab1}.
The jet structure (the angular energy and Lorentz factor profile) in each case is kept fixed, and results for minimal (without) lateral spreading are compared to the results with maximum lateral spreading.
Figure \ref{fig:ratios} shows the logarithm of the ratio of central parameter values for each parameter in our models, where individual radial spokes correspond to the labelled parameter ratio.
The radial scale indicates the value for the log ratio of the central parameter without spreading to the value with lateral spreading at the sound-speed; a ratio of 1 is indicated in the figure as a black dotted line.
In nearly all cases, the inclusion of lateral spreading changes the value of the parameter distribution central value.
We note that, for this data-set, the top-hat jet { structure} profile shows the most consistency between central parameter values, with only the ambient density, $n$, requiring a value a factor  $\sim3$ larger when spreading is neglected.
With the exception of the top-hat profile, the inferred inclination of each system and the core opening angle, is typically larger when lateral spreading is included.
For the top-hat jet, we see equal or consistent opening angles and all other parameters, with a marginally smaller inclination angle when lateral spreading is included. 
In terms of pre-peak temporal index, a smaller inclination angle for a given $\alpha$ is what is seen in Figure \ref{fig:alpha-0.3tp} for all jet structure models when lateral spreading is included for a fixed parameter system.
For structured jet systems, when fit to a real data-set, the variation in the other parameters, particularly the microphysical parameters $\varepsilon_B$ and $\varepsilon_e$, plus ambient density, $n$, which are required to fix the peak time and flux density, can absorb some of the expected change in inclination.
However, we note that when the ratio of inclination to core size is considered, $\iota/\theta_c$, a typically smaller value is seen for all jet structure models when lateral spreading is included, compared to without, see Figure \ref{fig:violin}.

\section{Discussion}\label{sec:disc}

We have made a comparison of the afterglows from identical parameter set structured jet models with and without the inclusion of lateral spreading in the calculation of the flux and observation time.
We have shown that, where everything else is the same, an afterglow model that includes lateral spreading will peak earlier and, for observers viewing the system from outside of the core opening angle, $\iota>\theta_c$, will have a rising lightcurve index that is higher, in all cases, than the comparable inclination and pre-peak time for an afterglow without lateral spreading.
{ This is a particularly important consideration when using afterglow modelling of the rise-index for a GW-EM counterpart to infer the systems inclination angle.}
Further, we have demonstrated how the inclusion, or omission, of lateral spreading in afterglow models can result in different preferred values for model free-parameters when fitting to observed data-sets.
This is in addition to the differences in parameters seen as a result of the jet { structure} choice.

Our lightcurves are calculated at an observed frequency of 3 GHz, at this frequency, the emission can be sensitive to synchrotron self-absorption (SSA).
For SSA flux post jet-break, the temporal behaviour of the lightcurve for an on-beam observer i.e. within the jet/core opening angle, $\iota<\theta_c$, or at the peak time and beyond for an initially off-axis observer, depends sensitively on the order of the synchrotron self-absorption frequency, $\nu_a$, the characteristic frequency, $\nu_m$, and the observer frequency, $\nu$.
Analytically, the pre jet-break afterglow lightcurve, where $\nu<\nu_a$, the flux will evolve with time as, $t^{1/2}$, for $\nu_a<\nu_m$, and $t^{5/4}$ where $\nu_m<\nu_a$.
Whereas, post jet break, these two conditions become, $t^0$ and $t^1$ for $\nu<\nu_a<\nu_m$ and $\nu<\nu_m<\nu_a$ respectively.
As $\nu_a<\nu$, and post jet-break, the flux is described with $t^{-1/3}$ and $t^{-p}$ for $\nu<\nu_m$ and $\nu>\nu_m$ respectively -- thus the SSA flux post jet-break may result in a flaring flux that will increase as $F\propto t$ until the passage of the characteristic frequency and/or the self-absorption frequency $\nu=\nu_a$.
Thus, variability at, or about the jet-break time for an on-axis observer, particularly at radio frequencies can be expected.
Whilst $\nu_a<\nu_m$, the self-absorption frequency will { remain constant} with observer time, $t$, while $\nu_m\propto t^{-3/2}$.
When $\nu_m<\nu_a$, then $\nu_a$ will begin to decline with time \citep{sari2002, gao2013}.
For an observer at a few $\sim\theta_c$, and where the afterglow is SSA at the jet-break time (in the on-axis frame), the peak of the afterglow (when $1/\Gamma \equiv (\iota-\theta_c)$) will be followed by a shallow decline phase before a break as $\nu_a$ crosses the observation band -- for very narrow or energetic jets, a flare { or brief increase in flux} before the final break may still be observed in SSA cases.
The exact SSA frequency depends sensitively on how it is estimated, a full discussion of SSA is beyond the scope of this current work -- our SSA follows that described in \cite{lamb2019b}, for an alternative estimate, see \cite{resmi2016}.

In Figure \ref{fig:alpha-0.3tp} we show that for an observer at $\iota>\theta_c$, the peak flux for the afterglow occurs at an earlier time where lateral spreading is included.
With our maximal spreading prescription, the peak time is a factor $\sim2$ earlier than the case without lateral spreading 
-- this factor represents the uncertainty { where the degree of lateral spreading is unknown}. 
{ Whereas} this difference in the peak time for jets with identical parameters, note that the peak flux level, $F_p$, is unaffected by the inclusion or not of spreading, will result in different parameter sets for identical jet models dependent on whether spreading is considered or not.
Figure \ref{fig:alpha-0.3tp} also shows the rise index at a fraction, 0.3, of the peak time.
The same rise index is achieved at smaller inclinations where spreading is included.
Naively, this would imply that the inclusion of spreading in a model would result in smaller inferred inclination angles from the same data-set when compared to a non-spreading model, however, by considering the difference in the peak time, then we can see that the other free parameters will come into play as $t_p \propto (E/n)^{1/3}$ and $F_p \propto E n^{(p+1)/4} \varepsilon_B^{(p+1)/4} \varepsilon_e^{p-1}$.

In fitting four different jet { structure} models, each with and without spreading, to the afterglow data-set for GRB\,170817A via an MCMC we get distributions for the model free parameters, see Figure \ref{fig:GW170817} for a sample of the posterior distribution lightcurves.
The central and 16th and 84th percentile ranges for the parameters are listed in Table \ref{tab1}.
Other than through the post-peak decline, the difference in spreading or non-spreading models is not immediately evident from these lightcurves.
The difference in late-time decline is noted in \cite{troja2018} and \cite{lamb2018a} in reference to breaking the model degeneracy in relation to a relativistic core dominated jet, or a quasi-spherical stratified cocoon as the origin for the slow rising afterglow lightcurve of GW170817 pre-peak at $\sim150$\,days.
Here, we show that besides the choice of jet structure \citep[e.g.][]{ryan2020}, the inclusion or omission of lateral spreading will also affect the parameter values.
Figure \ref{fig:ratios} shows the log ratio of the central parameter values for each model.
The parameters with the biggest difference for the same structure profile are the microphysical parameters, $\varepsilon_B$ and $\varepsilon_e$.
Of our fiducial jet structures, the powerlaw { structure} has the largest difference in parameter values, and the top-hat { structure} has the smallest difference with only the ambient density showing a reasonable difference at a factor $\sim3$.

In terms of the inclination, which is the parameter of main interest in this study, we note that all structures, except the top-hat, have larger preferred inclination angles $\iota$ when spreading is included.
For our sets, this should be considered in ratio with the preferred core angle, $\theta_c$.
For GW170817, limits on the core size and the relativistic motion of the jet core was constrained via Very Long Baseline Interferometry (VLBI) \citep{ghirlanda2019, mooley2018}.
The VLBI measurements can be used to constrain the ratio of inclination to the core size, with a preferred $\iota/\theta_c\sim 5\pm1$ \citep{nakar2021}.
For values of $\iota/\theta_c\lesssim 5$, \cite{fernandez2021} showed that synthetic image modelling can determine the true inclination angle with an uncertainty $\lesssim10\%$ on $\iota$.

In Figure \ref{fig:violin} we show the posterior distributions for $\iota/\theta_c$ for each model { fit to the data}, both with and without lateral spreading.
We note that, although there can be significant overlap, where lateral spreading is included the central value for the ratio $\iota/\theta_c$ is typically smaller than the same model without lateral spreading.
Of all our models fit to the data\footnote{ We note that the powerlaw jet { structure} that we use follows the definition in \cite{lamb2017} and is different to the powerlaw { structure} profiles employed by \cite{hotokezaka2019} and \cite{ghirlanda2019}.}, only the Gaussian structure jet without lateral spreading has the central and 16th -- 84th percentile limits for $\iota/\theta_c$ within the expected 4 -- 6.
Inspection of the lightcurve for this model in Figure \ref{fig:GW170817} shows good agreement with the data, including the early and late-time X-ray data as presented in \citep{troja2021}, where we use the late data binned so that a general declining trend is supported (and consistent with the expectation from a GRB-like afterglow).
With this non-spreading Gaussian jet { structure}, there is no evidence of any excess in late-time radio or X-ray data, with an inclination of $\iota=0.29\pm0.07$ rad, or $\sim16.7\pm4.0$ degrees.
By considering the $\iota/\theta_c$ ratios, both of the two-component models, with and without spreading have a significant overlap with the VLBI constraints.
Similar to the non-spreading Gaussian model, both of the two-component models give good agreement with the observed late-time radio and X-ray decline, and an inclination of $\iota=0.23\pm0.01$ rad ($13.2\pm0.6$ degrees) without spreading and $\iota=0.32^{+0.22}_{-0.02}$ rad ($18.3^{+12.6}_{-1.3}$ degrees) with lateral spreading.

The future detection of GW-EM afterglows from core dominated relativistic jets could be used to help constrain the degree of lateral spreading in GRB afterglows, however, the effects of jet structure choice would need to be better understood or the physically expected jet { structure} for short GRB jets found \citep[e.g.][and Nativi et al. in prep]{salafia2020, murguiaberthier2021, wang2021}.

\section{Conclusions}\label{sec:conc}

We have made comparisons of { GRB and GW-EM} afterglows modelled with and without lateral spreading as seen by observers at various inclinations. We find that the inclusion of lateral spread in the afterglow lightcurve estimation can affect the rise index pre-peak, and the peak time, where all other parameters are fixed.
When fitting observed data-sets with an afterglow model, the inferred parameters depend sensitively on the choice of jet structure but additionally on the whether lateral spreading is included or not.
For all jet { structure} profiles tested we find that the ratio, $\iota/\theta_c$ is typically smaller where lateral spreading is included in the afterglow model.
{ Although the ratio is smaller where lateral spreading is included, the}
individual values of the inclination and core angle may be larger in either the spreading or the non-spreading cases.
We find that for structured jets other than a top-hat { structure}, $\theta_c$ and $\iota$ are { typically }larger where spreading is included.
The peak flux and peak time can depend sensitively on other free parameters in the afterglow model, and there is no clear trend for larger or smaller values for any of these in all structure profiles.
The ratio of the parameter values without and with spreading depends more sensitively on the choice of jet structure.

We caution that, where estimates of the inclination, $\iota$ are made using afterglow rise index fitting for a given jet structure profile (useful for adding additional constraints on the Hubble parameter \citep{mastrogiovanni2020a}), then these results should consider complimentary fitting of radio images of superluminal motion, if available, \citep[e.g.][]{fernandez2021}, or fold in the uncertainty on inclination from the inclusion or not of a spreading prescription (which can be unique to the structure model used).
We find a factor $\lesssim2$ on the inclination angle, $\iota$, from data fits with the same jet structure profiles, dependent on whether spreading is included or not.

\vspace{6pt} 



\authorcontributions{Conceptualization, GPL, MH; methodology, GPL; validation, JJF; writing---original draft preparation, GPL; writing---review and editing, FH, AKHK, NRT, ISH, SS, JV; visualization, GPL, FH. All authors have read and agreed to the published version of the manuscript.}

\funding{GPL is supported by STFC grant ST/S000453/1.}

\institutionalreview{Not applicable.}

\informedconsent{Not applicable.}

\dataavailability{Data available on reasonable request to the authors.} 

\acknowledgments{GPL thanks Gordon Stewart for discussions that motivated this study. The authors thank Laurence Datrier, and Michael J. Williams for useful discussions.}

\conflictsofinterest{The authors declare no conflict of interest.} 



\abbreviations{The following abbreviations are used in this manuscript:\\

\noindent 
\begin{tabular}{@{}ll}
GRB & Gamma ray burst\\
GW & Gravitational wave\\
EM & Electromagnetic\\
GW-EM & Gravitational waves and electromagnetic emission\\
TH & Top-hat\\
G & Gaussian\\
2C & Two-component\\
PL & Powerlaw\\
MCMC & Markov chain monte carlo\\
w & With\\
wo & Without\\
VLBI & Very long baseline interferometry\\
GHz & Giga-Hertz\\
STFC & Science technology and facilities council
\end{tabular}}


\reftitle{References}


\externalbibliography{yes}
\bibliography{ms.bib}


%


\end{paracol}

\end{document}